\pdfoutput=1

\documentclass[11pt]{article}

\usepackage[preprint]{acl_latex}

\usepackage{times}
\usepackage{latexsym}
\usepackage{multirow}

\usepackage[T1]{fontenc}

\usepackage[utf8]{inputenc}

\usepackage{microtype}

\usepackage{inconsolata}

\usepackage{graphicx}
\graphicspath{ {./resources/} }

\title{\textbf{\textsc{KI4Demokratie}}: An AI-Based Platform for Monitoring and Fostering Democratic Discourse}

\author{
 \textbf{Rudy Alexandro Garrido Veliz\textsuperscript{1}},
 \textbf{Till Nikolaus Schaland\textsuperscript{1}},
 \textbf{Simon Bergmoser\textsuperscript{2}},
 \\
 \textbf{Florian Horwege\textsuperscript{1}},
 \textbf{Somya Bansal\textsuperscript{3}},
 \textbf{Ritesh Nahar\textsuperscript{3}},
 \\
 \textbf{Martin Semmann\textsuperscript{1}},
 \textbf{Jörg Forthmann \textsuperscript{4}},
 \textbf{Seid Muhie Yimam \textsuperscript{1}}
\\
\\
 \textsuperscript{1}Universtät Hamburg,
 \textsuperscript{2}NORDAKADEMIE,
 \textsuperscript{3}Zirkel Technologies GmbH,
\\
 \textsuperscript{4}FAKTENKONTOR,
\\
 \small{
   \textbf{Correspondence:} \href{mailto:rudy.garrido.veliz@uni-hamburg.de}{rudy.garrido.veliz@uni-hamburg.de} \& \href{mailto:seid.muhie.yimam@uni-hamburg.de}{seid.muhie.yimam@uni-hamburg.de}
 }
}

\begin{document}
\maketitle
\begin{abstract}


Social media increasingly fuel extremism, especially right-wing extremism, and enable the rapid spread of antidemocratic narratives. Although AI and data science are often leveraged to manipulate political opinion, there is a critical need for tools that support effective monitoring without infringing on freedom of expression. We present \textbf{\textsc{KI4Demokratie}}, an AI-based platform that assists journalists, researchers, and policymakers in monitoring right-wing discourse that may undermine democratic values. \textbf{\textsc{KI4Demokratie}} applies machine learning models to a large-scale German online data gathered on a daily basis, providing a comprehensive view of trends in the German digital sphere. Early analysis reveals both the complexity of tracking organized extremist behavior and the promise of our integrated approach, especially during key events.

\end{abstract}

\section{Introduction}

Democracy is currently under threat globally and especially across several European nations due to a significant rise in extremist ideologies, particularly from right-wing factions. These groups have used the principle of freedom of speech as a means to spread polarized and often extremist narratives in social networks and beyond, leading to increased societal divisions \cite{Aktas2024, Pantucci2024}. The presence of right-wing political actors and their influence in digital spaces such as social media is undeniable, amplifying their ideologies \cite{Serrano2019}. This phenomenon is further complicated by the strategic use of artificial intelligence and data science to manipulate public opinion and undermine democratic processes \cite{Heawood2018,Khanal2025,Riedl2024}.

In response to these challenges, we propose the \textbf{\textsc{KI4Demokratie}} project, a pioneering initiative aimed at leveraging artificial intelligence to counteract the influence of antidemocratic forces. This project focuses on developing a comprehensive dashboard that helps journalists, politicians, policy makers, and researchers analyze data from social media platforms and online news outlets. The ultimate goal is to preserve democracy and to create safer social media and news portals, fostering the coexistence of diverse viewpoints and mitigating community fragmentation.

The \textbf{\textsc{KI4Demokratie}} dashboard is composed of three core components: 1) Sentiment and Hate Speech Analysis Module: This component employs advanced AI models to automatically analyze sentiment and hate speech from various data sources. It provides information on the distribution of emotional and hostile content, highlighting prevailing trends in societal polarization \cite{Kirill2022, Yimam2024}.
2) Network Analysis Module: This feature elucidates potential connections between political entities through sophisticated network graphs, providing insight into structural relationships within political ecosystems and the dissemination pathways of ideologies \cite{Philip2017,Chua03042025}.
3) Topic Extraction and Visualization Module: This component focuses on identifying and visualizing key topics from the collected data, enabling users to derive meaningful insights and enhance their understanding of evolving political narratives \cite{xu_setting_2020}.

The research framework underpinning this project is guided by several key questions: \textbf{RQ1}: What are the predominant themes and narratives disseminated by extremist groups on social media? \textbf{RQ2}: How effective are AI-driven methodologies in identifying and countering extremist narratives and misinformation? \textbf{RQ3}: In what ways can \textbf{\textsc{KI4Demokratie}} contribute to the development of safer online spaces and the promotion of integrated communities?
Our contributions are multifold: First, we introduce a novel analytical tool, the \textbf{\textsc{KI4Demokratie}} dashboard, which integrates sentiment analysis, network mapping, and topic modeling to provide a holistic view of political discourse. Second, we demonstrate the application of artificial intelligence in monitoring and countering extremism, thereby offering a blueprint for future initiatives in media oversight and democratic safeguarding. Third, we provide empirical insights into online extremist rhetoric dynamics, contributing to academic knowledge and practical policy-making efforts.
These contributions have several implications: They offer a scalable model for monitoring political discourse across various digital platforms, assist stakeholders in developing informed strategies to counteract polarization, and ultimately, serve as a foundation for fostering more constructive and cohesive public dialogue.


\begin{figure}[t!]
\centering
\includegraphics[width=\columnwidth]{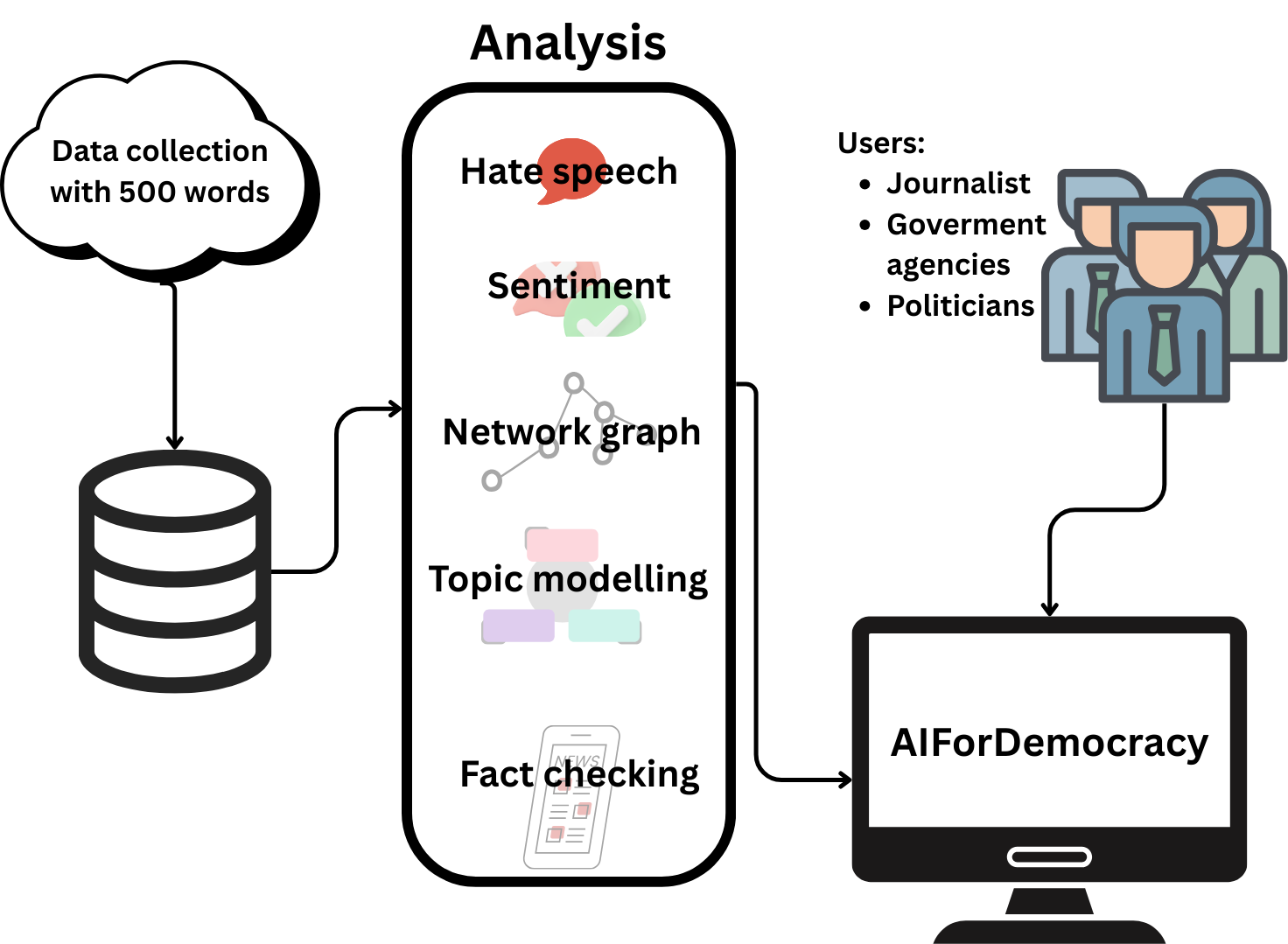}
\caption{Complete pipeline of the \textbf{\textsc{KI4Demokratie}} project.}
	\label{fig:pipeline}
\end{figure}

\section{Related Work}

The rapid digital spread of antidemocratic ideologies and the organized mobilization of right-wing groups present an escalating challenge for democratic institutions globally. Foundational studies have systematically delineated the mechanisms of digital populism \citep{bartlett2011newface, mudde2019far}, while research demonstrates how alternative platforms and digital media ecosystems shape and aid radical right agendas \citep{Conway2019}. Advances in automated hate speech and sentiment detection have improved coverage for diverse languages, including significant contributions to model development and resource creation \citep{schmidt2017survey, vidgen2019abusive, wiegand2018abusive}, with benchmark datasets such as GermEval supporting evaluation in the German context \citep{germeval2018}. 

Extremist activity is increasingly fragmented across platforms, requiring robust and multifaceted network analysis. Methodologies for tracing actor relationships and information flows are detailed in prior work \citep{morstatter2013twitterapi, zannettou2019web, ribeiro2021youtube}, yet truly integrative tools that capture influence across multiple platforms are still rare. Similarly, topic modeling, from LDA \citep{blei2003lda} to BERTopic \citep{grootendorst2022bertopic}, as discussed in applied communication contexts \citep{jacobi2016news}, enables the extraction of latent themes, but dynamic narrative tracking and combined approaches with network or sentiment analytics remain exceptions. Efforts to counter disinformation, such as AI-supported fact-checking \citep{nakov2021factcheck} and analyses of computational propaganda \citep{howard2016propaganda}, have often been limited by a lack of integrated, practitioner-friendly interfaces.

Addressing these gaps, our \textbf{\textsc{KI4Demokratie}} initiative pioneers an integrated solution that combines sentiment analysis, hate speech detection, dynamic topic modeling, and advanced network analysis in an interactive dashboard. This holistic, cross-methodological approach delivers urgently needed scalable tools for researchers, journalists, and policymakers working to protect and strengthen democratic discourse in the digital age.

\section{Approaches}

The data is obtained daily from varied sources, based on a distilled set of 500 keywords from human experts. The obtained posts go through text classification for sentiment, hate speech, claim identification, and fact checking. In addition to this, a topic modeling analysis is performed and a network graph is created. These steps are detailed in the following sections.

\subsection{Analysis Dashboard}

The analytical dashboard offers an interactive visualization of sentiment and hate speech trends derived from online content. Developed using Dash and Plotly, the dashboard would integrate the outputs of two types of text classification, sentiment analysis, and hate speech, using small language models (SLMs). For sentiment analysis, the system uses the TimeLM 
model \citep{Loureiro2022}, which computes compound sentiment scores for each entry, classified as positive, negative, or neutral. 
For the hate speech classification, we employ the SLM  model, LFTW R4 Target 
\citep{Vidgen2020}, fine-tuned on hate speech data. Each text input is tokenized and passed through the model, which outputs logits. These logits are converted into binary labels using argmax, identifying whether a text contains hate or normal speech. The sentiment and hate speech models used are state-of-the-art and best-performing in English, and we apply translation to German to utilize them.
%
%
The dashboard empowers users to analyze extremist language and online narratives by offering dynamic visualizations of sentiment and hate speech trends over time and across platforms, enabling targeted exploration of discourse patterns, shifts, and spikes related to political events or media campaigns. 

\subsection{Topic Modeling}\label{subsec:approach_topic_model}
In order to uncover predominant themes and narratives by online news and right-wing groups on social media, we employ topic modeling to identify underlying semantic structures.
The topic extraction and visualization component utilizes BERTopic \citep{grootendorst2022bertopic}, a robust topic modeling technique that relies on transformer-based embeddings to uncover latent themes in large corpora of text and produces interpretable topics by applying class-based TF-IDF.
For the generation of document representations, the pre-trained sentence-transformer, Sentence-BERT \citep{reimers-2019-sentence-bert}, is applied.
The subsequent dimensionality reduction step in BERTopic was performed using UMAP \citep{mcinnes2018umap-software}, followed by HDBSCAN \citep{mcinnes2017hdbscan} as the clustering algorithm.
The selection of parameters aimed to mitigate the identification of spurious, unrelated clusters while avoiding the aggregation of distinct themes into overly broad topics and to prevent the fragmentation of coherent topics into excessively fine-grained clusters.
Specifically, we apply BERTopic to perform dynamic topic modeling on a daily basis to model the evolving political narratives and themes.
By tracking the frequency and content of identified topics daily, journalists and politicians can pinpoint the predominant themes at any given moment in time, observe their trend, and link these dynamics to external resources and real-world events, see examples in Appendix \ref{sec:dtm_appendix}.

\subsection{Network Graph}
 Narratives and their influence online extend beyond a single social network. Finding influential actors on multiple social media networks, news articles, blogs, and the Internet in general is not trivial. Some articles find that the heavy reliance of research on social network graphs is based on access to the Twitter/X API \citep{morstatter2013twitterapi}. The proposed network extraction method and centrality calculation approach aims to address this issue.
The available interaction data provides the author of a post or news snippet and the URL at which this interaction occurs. Both entities can be extracted and transformed into the origins of an edge in a social network graph. The author is extracted directly, and the URL is extracted as a user profile, in which case it is matched to the entity of the author or the organization hosting the interaction. The extraction method then transforms and extracts data from the content of the interaction itself to select what kind of edge in the graph should be attributed to this interaction and what the target node should be.

One way of extracting the target node is by checking for tagged users, for example @-mentions, using regular expressions. Another method is by using named entity recognition (NER). The method of extraction determines which of the two labels the edge between the source and the target node has. The entities labeled are the targets of edges labeled \textit{intentional}, while entities extracted using NER are the targets of the edges labeled \textit{ inferred}. All edges are weighted by the frequency of their occurrence.
Some entities, while not very active themselves, are mentioned with a high frequency and are therefore important for the understanding of influential actors in the network. To include this aspect of influence, the method also extracts a third type of edge, \textit{passive mutual}, which is undirected. It connects nodes that are mentioned within the same content of an interaction by a 3rd party \citep{rabigerFrameworkValidatingMerit2015}

The network graph also includes hashtags and the modeled topics as a separate node label. Hashtags are the targets of the edge labeled \textit{intentional}, while topics extracted using the topic modeling approach are targets of the edge labeled \textit{inferred}. 
Eigenvector centrality \citep{newmanNetworksIntroduction2016} would be used to scale the nodes in a view of the graph to give end-users orientation to identify influential nodes. Nodes should also be filtered based on their frequency of occurrence in the data. 

\subsection{Fact Checking}
In order to assess the factuality of a text, a multi-stage framework has been proposed \cite{Vlachos.2014, Guo.2022}. The stages are (i) claim detection, (ii) evidence retrieval, and (iii) claim verification. We developed a prototype adopting this framework to evaluate its feasibility in connection with a results dashboard.
For the stage of claim detection, in which the aim is to identify a check-worthy claim within a text and to extract this claim, one-shot prompting with GPT-3.5 \cite{brown2020languagemodelsfewshotlearners} was utilized. The model was supplied with a system prompt (see Appendix \ref{app:prompt}), which contained the example, and a user prompt, which contained the post as well as the author, the author party, and the date as context information.
The extracted claim was used for the next stage of evidence retrieval. Evidence retrieval aims to find sources in connection to the stated claim from Google News search. 
For the generation of the search query, one shot prompting with GPT-3.5 was used again. As in the previous setup, the system prompt contained an example, while the user prompt contained the previous reply and context information. The first 3 results of the Google News search were used as knowledge sources and summarized by GPT-3.5. The concatenated summaries were then passed to the next stage of verdict prediction.
For the verdict prediction, GPT-3.5 was supplied with the claim, context information and context information. In a zero-shot prompting setup, it was tasked to give a truthfulness score out of five categories (\emph{False}, \emph{Mostly false}, \emph{Half true}, \emph{Mostly true}, or \emph{True}) and give a reason of at most 2 sentences, and truthfulness scores were displayed on the dashboard. For each channel, the number of claims in each of the five categories was displayed in a bar chart, giving an overview of how often each author published claims that were supported/not supported by evidence. This planned approach is experimental; among the other considered ones is the one by \citet{Sevgili2024} stands out, for their streamlined approach. 

\section{Evaluation}
To evaluate our approach, 100 random posts across various dates were extracted and given to four different German native speakers. The annotators were asked to categorize the posts regarding sentiment (positive, negative, or neutral) and whether they contain hate speech. Where the four annotators were evenly split, a fifth native speaker acted as a tie-breaker, and the gold label for each post was decided by majority vote. The Kappa score was around 0.43 for both tasks. This shows how challenging even for native speakers to coincide in these matters; and therefore, interpretation of experts, for example, journalists or researchers, is needed to effectively assess the behavior of the right-wing movement. Applying the LFTW R4 Target model for hate speech, the TimeLM model for sentiment analysis, and GPT-4o-mini to the gold label dataset shows moderate performance, with GPT-4o-mini achieving the best results overall (see Table \ref{table:model-metrics}).
%
%

\begin{table}[ht]
\centering
\small
\begin{tabular}{|l|l|c|c|c|}
\hline
\textbf{Task}      & \textbf{Model}        & \textbf{Precision} & \textbf{Recall} & \textbf{F1} \\
\hline
\multirow{2}{*}{Hate speech}
    & LTFTW            & 0.42 & 0.85 & 0.56 \\
\cline{2-5}
    & GPT     & 0.53 & 0.63 & 0.58 \\
\hline
\multirow{2}{*}{Sentiment}
    & TimeLM            & 0.77 & 0.72 & 0.74 \\
\cline{2-5}
    & GPT      & 0.86 & 0.82 & 0.81 \\
\hline
\end{tabular}
\caption{Hate speech and sentiment analysis using SLM (LFTW R4 Target and TimeLM respectively) and LLM (GPT 4o-mini) models.}
\label{table:model-metrics}
\end{table}
To illustrate the capabilities of the Dynamic Topic Modeling approach described in Section \ref{subsec:approach_topic_model}, we present two example visualizations derived from the dashboard.
Specifically, we focused on the period around the \textbf{Aschaffenburg}\footnote{\url{https://www.dw.com/en/germany-Aschaffenburg-knife-attack-details/a-71372024}} incident in January 2025.
Figure \ref{fig:dtm_general_topics} and Figure \ref{fig:dtm_focus_topics} in Appendix \ref{sec:dtm_appendix} describe the temporal evolution of selected topics during this timeframe, clearly demonstrating the incident's impact on the social media discourse and the increase in hate speech-related themes.
Furthermore, the visualization also demonstrates the subsequent rise in topics related to migration policy, correlating with a proposal introduced by a conservative political party.
These examples underline the capability of the models to pinpoint predominant themes and narratives at specific points in time and link their fluctuations to real-world events and news.


\section{Conclusion}
We have carefully developed a strategy to effectively leverage AI against right-wing anti-democratic movements. The components of the dashboard could support journalists, researchers, and policymakers in identifying and responding to problematic narratives and themes in the digital sphere. In a preliminary qualitative analysis, we have found a cohesive movement of the right-wing discourse when troubling situations arise, for example, the \textbf{Aschaffenburg} incident that led to an increase in hate speech. Our work, as it is still in progress, is trying to discover the best ways of using AI to support the resilience of democratic societies. 

\section{Limitations}  
There are multiple limitations that can be faced when creating such a complex tool. The sheer amount of data and the capabilities for analysis are one of the biggest ones, while also having to mind the resource component. 

Another limitation is how hard it is to get direct feedback from the possible users in order to improve it. These users are often in high demand, so it is a challenge to progress at a good pace while satisfying the possible requirements they have.

The accessibility of language models trained in the German language is also a limitation, considering that the ones found are often underdeveloped. The research community must strive for it. 

However, the advance of multilingual language models offers fine-tuning opportunities for specific tasks. A possible improvement for the fact-checking framework would be fine-tuned models for the specific subtasks. A current work in progress is the fine-tuning of a large language model to identify a claim in a text and extract it in a normalized form, as suggested by \citet{
sundriyal-etal-2023-chaos}.     
\section*{Ethical Considerations} 
The main consideration taken is that our work does not assume the political ideology, or all the social expectations that may come with it, of individuals. Rather, it only deals with the texts obtained to get a grasp of the right-wing discourse. We focus exclusively on analyzing publicly shared speech, not individuals, and emphasize transparency and academic rigor in countering anti-democratic actions.

It is also understood how important it is, for a set of tools like the one we intend to create, to keep access to it only for the desired individuals. A curated ingress to the dashboard will be created. Access to this tool is strictly controlled to prevent misuse for anti-democratic purposes. 


\bibliography{acl_latex}

\appendix

\section{Fact Checking}
\label{app:prompt}
The fact-checking prototype used three \textit{one-shot} prompting configurations and one \textit{zero-shot} setup. All prompting was performed with OpenAI's GPT-3.5 model (\texttt{gpt-3.5-turbo}).

\subsection*{Claim Extraction}

To extract fact-checkable claims from text, the system was prompted as follows:

\textbf{System prompt:}
\begin{quote}
\textit{You are a fact expert. Your task is to extract the statements from a given text for which the truthfulness can be verified by publicly available knowledge. \\
Ignore information about personal events, subjective opinions, and statements about upcoming events. \\
Reformulate the statements into direct statements. Include the date, author, and author's party in your formulation if this is necessary for fact-checking the statement.\\
Example: Text: "You can see the bad mood among all other parties."; Author: "Christian Lindner from the FDP party"; Date: 1.2.2023; Direct statement: "In February 2023, there is a bad mood among all parties except the FDP."\\
Return only the direct statements as a JSON object. Name these \texttt{"statements"}.}
\end{quote}

\textbf{User prompt template:}
\begin{quote}
\textit{Extract the statements that can be factually verified with public knowledge from the following text by \{author\} from the party \{author\_party\} from \{date\}: \{query\}.}
\end{quote}

\subsection*{Query Generation}

To rephrase extracted claims into optimized Google search queries, the following prompts were used:

\textbf{System prompt:}
\begin{quote}
\textit{You are a Google search expert. Your task is to create an optimized Google search query from a statement. \\
Example: Input: "The capital of Germany is Berlin." Optimized Google search query: "Capital of Germany"}
\end{quote}

\textbf{User prompt template:}
\begin{quote}
\textit{Convert the following statement by \{author\} from \{date\} into a Google search query: \{query\}.}
\end{quote}

\subsection*{Verdict Prediction (Zero-Shot)}

The zero-shot setup for predicting truthfulness (verdicts) used the following configuration:

\textbf{System prompt:}
\begin{quote}
\textit{You are a fact-checking expert. Your job is to check the truth of statements that are given to you. You are provided with contextual knowledge for this purpose. \\
Only give your answer in JSON format with the fields \texttt{"Truthfulness"} and \texttt{"Reason"}. \\
Select exactly one of the following categories for the \texttt{"Truthfulness"} field: \texttt{"False"}, \texttt{"Mostly false"}, \texttt{"Half true"}, \texttt{"Mostly true"}, or \texttt{"True"}. \\
The \texttt{"Reason"} field must not be longer than two sentences.}
\end{quote}

\textbf{User prompt template:}
\begin{quote}
\textit{Check the following statement by \{author\} from the party \{author\_party\} from \{date\}: \{claim\}. \\
Context: \{grounding\_context\}}
\end{quote}

\section{Dynamic Topic Modeling}\label{sec:dtm_appendix}
\begin{figure*}[h!]
    \centering
    \includegraphics[width=\linewidth]{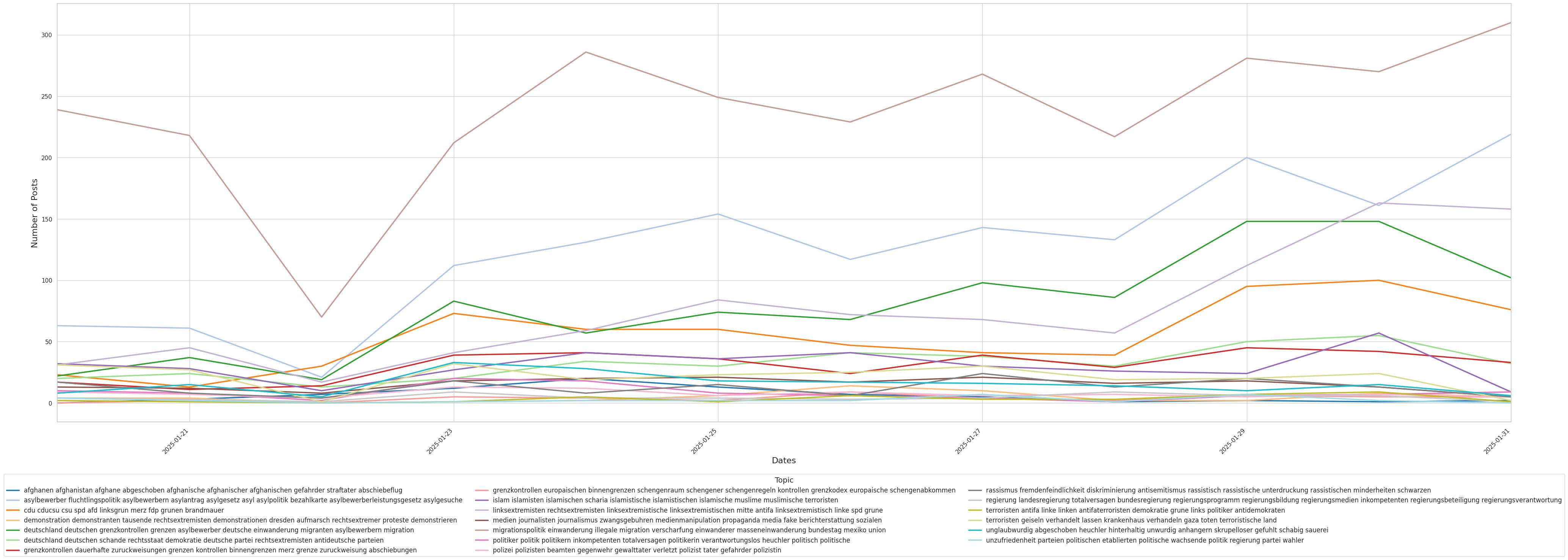}
    \caption{Temporal evolution of a wide range of political and extremist discourse topics on social media from January 20 to February to January 31, 2025, covering the period around the \textbf{Aschaffenburg} incident and subsequent proposal by a conservative political party.}
    \label{fig:dtm_general_topics}
\end{figure*}

\subsection{Focus Topics}
\begin{figure*}[h!]
    \centering
    \includegraphics[width=\linewidth]{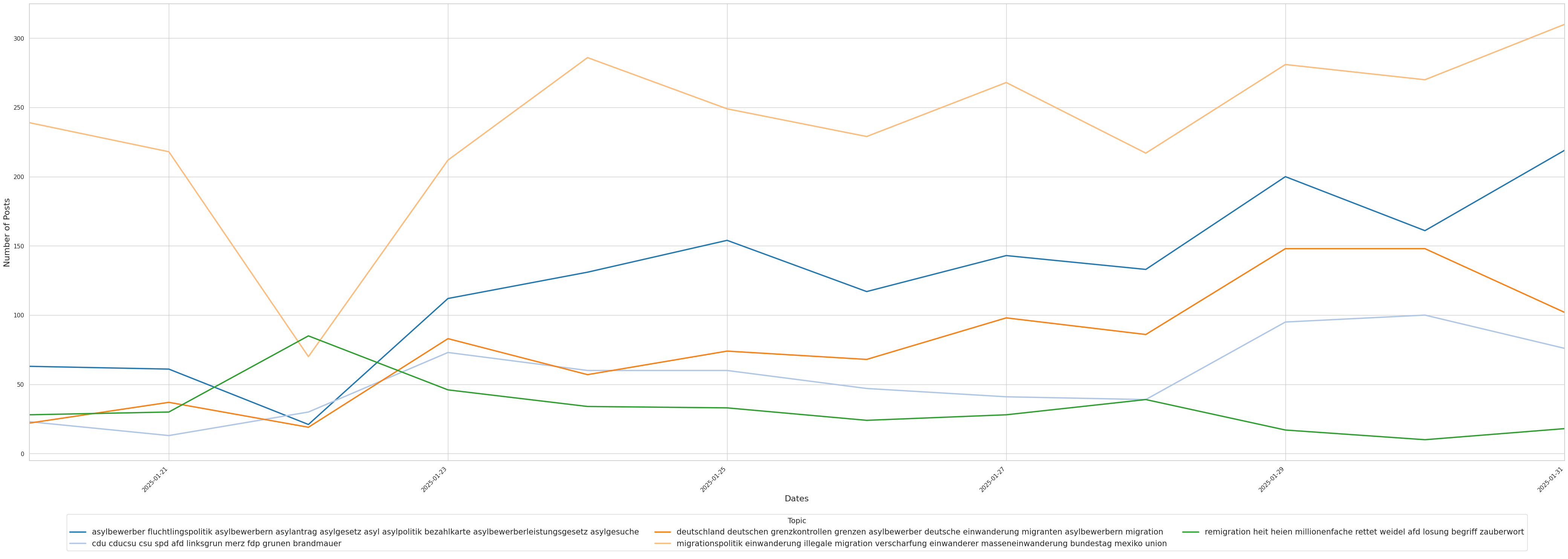}
    \caption{Evolution of key political discourse topics on social media from two days before the incident in \textbf{Aschaffenburg} (January 22, 2025) until January 31, 2025, coinciding with the joint proposal by a conservative political party. The plot highlights trends in  discourse related to migration, border control, asylum policy, and right-wing narratives such as "Remigration".}
    \label{fig:dtm_focus_topics}
\end{figure*}

\end{document}